# FEATURES OF THE INTRA-MASS BUOYANCY.

Kochin A.V.

Central Aerological Observatory

3 Pervomayskaya str., Dolgoprudny, Moscow region, 141707 Russia

Email: amarl@mail.ru

**Abstract**

The buoyancy force is the cause of ordered vertical movements in the atmosphere, therefore, the analysis of the causes and conditions of its formation is important not only for the formation of convective clouds, but also for understanding all atmospheric transport processes. Due to the absence of rigid boundaries inside the gas, a horizontal pressure gradient in a static state cannot exist in the allocated volume. The pressure inside the allocated volume with a different density is equal to the external pressure and the intra-mass buoyancy force, according to its definition, is formally zero. The observed force of intra-mass buoyancy arises due to the difference in vertical pressure gradients in media with different densities. In this case, the buoyancy force is volumetric, and its value corresponds to generally accepted ratios.

1. Introduction

The buoyancy force is the cause of ordered vertical movements in the atmosphere. The most widely known convective instability is caused by the action of the buoyancy force, the magnitude of which depends on the vertical stratification of temperature. Convective instability leads to the formation of clouds, which contributes to the entry of water vapor from the underlying surface into the atmosphere. In the future, water vapor in the process of turbulent mixing and transport by ordered air (horizontal and vertical) flows enters the upper layers of the atmosphere and causes precipitation during its condensation. Analysis of the causes and conditions of buoyancy force formation is important not only for the formation of convective clouds, but also for understanding all atmospheric transport processes.

2. Hydrostatic equilibrium.

Hydrostatic equilibrium characterizes the stationary state of a gas in a gravitational field (Holton 2004, Lima F M S. 2015, Giancarlo Cavazzini 2018). The balance of pressure forces is shown in Fig.1.

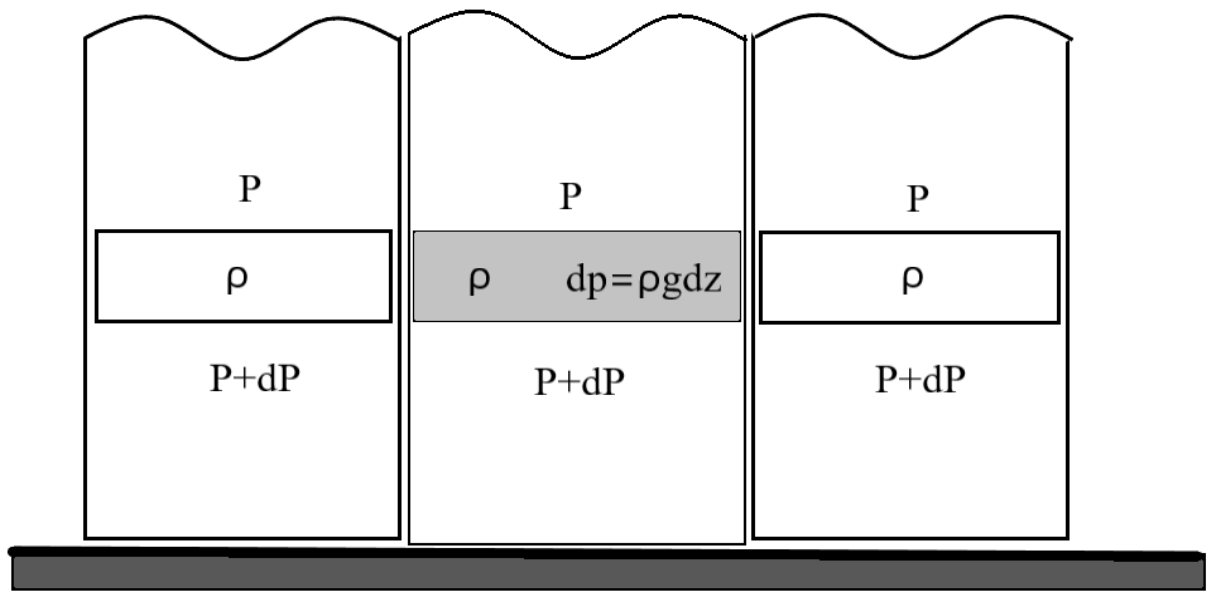

Fig.1. The balance of pressure forces in hydrostatic equilibrium. In a single column centered at a certain height, the pressure is P at a gas density of p. When the height decreases by dz, the pressure increases by dp= pgdz due to the weight of the gas in the dz layer. There is no horizontal pressure difference, because the density p is assumed to be unchanged horizontally.

### 3. Buoyancy and buoyant force

The buoyant force is defined as the resultant of all pressure forces on the outer surface of the body. The substance inside the volume creates pressure on the surface from the inside. The difference between the buoyant force and the pressure from the substance inside the surface determines the buoyancy. Since weight is equal to the product of density by volume, with different densities inside the volume and in the surrounding space, buoyancy can be both positive and negative. In the case of a body with rigid boundaries, the formal definition of buoyancy does not cause difficulties due to the known surface over which the integral of pressure is calculated. In the case of the intra-mass buoyant force, there are no rigid boundaries anymore. The difference in the formation of the buoyancy force is shown in Fig. 2.

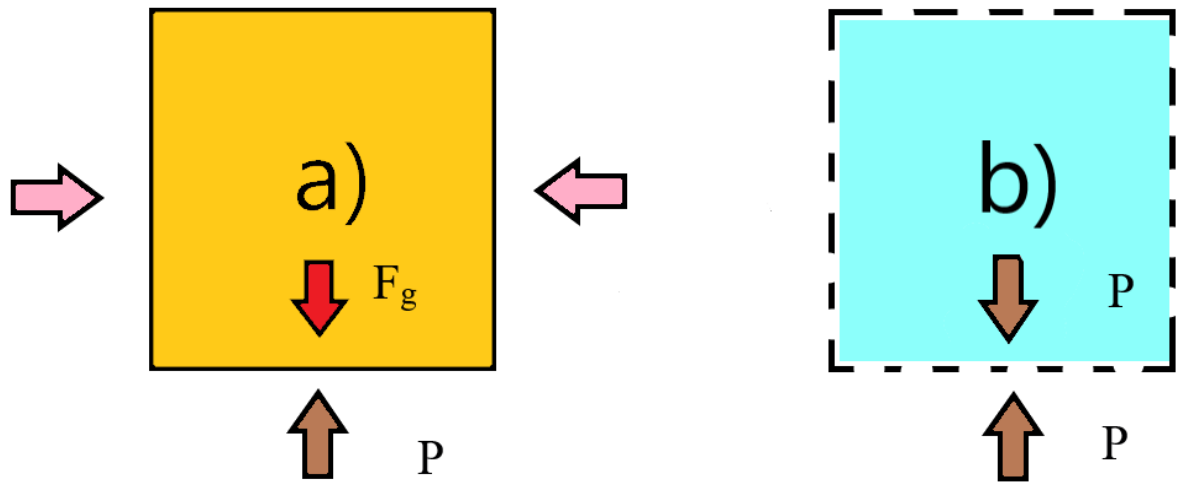

Fig. 2. Formation of the buoyancy force in the case of a body in the form of a cube with rigid boundaries (a) and a volume of gas (b) of the same shape with a different density.

In the case of a body with rigid boundaries, the pressure on the outer sides balances each other, so their resultant is zero (Fig. 2a). The lower part of the body surface is affected by external gas pressure (shown by the brown arrow), and the inner part is affected by body weight (shown by the red arrow). The difference of these forces determines the buoyancy force. In a gas, the horizontal pressure difference is zero in the static state (Fig. 2b). Since the difference is zero at any height, it is zero at the level of the base of the volume. Therefore, the external pressure on the lower part of the surface (shown by the brown arrow) is equal to the internal pressure (also shown by the brown arrow), and there is no pressure difference. According to the definition of the buoyancy force, the buoyancy force is identically zero regardless of the density of the gas inside the volume, because everywhere the external pressure and the internal pressure are equal to each other. Nevertheless, the buoyancy force in gases is not zero, therefore, the determination of the buoyancy force requires clarification.

### 4. Specific buoyancy force in gas

The buoyancy force is defined as the pressure difference between the outside and inside of the allocated volume. There is no difference between external and internal pressures for the allocated volume in the gas in a static state, therefore, the calculation must be performed using a different algorithm. Figure 3 shows a model for the formation of buoyancy forces in gases.

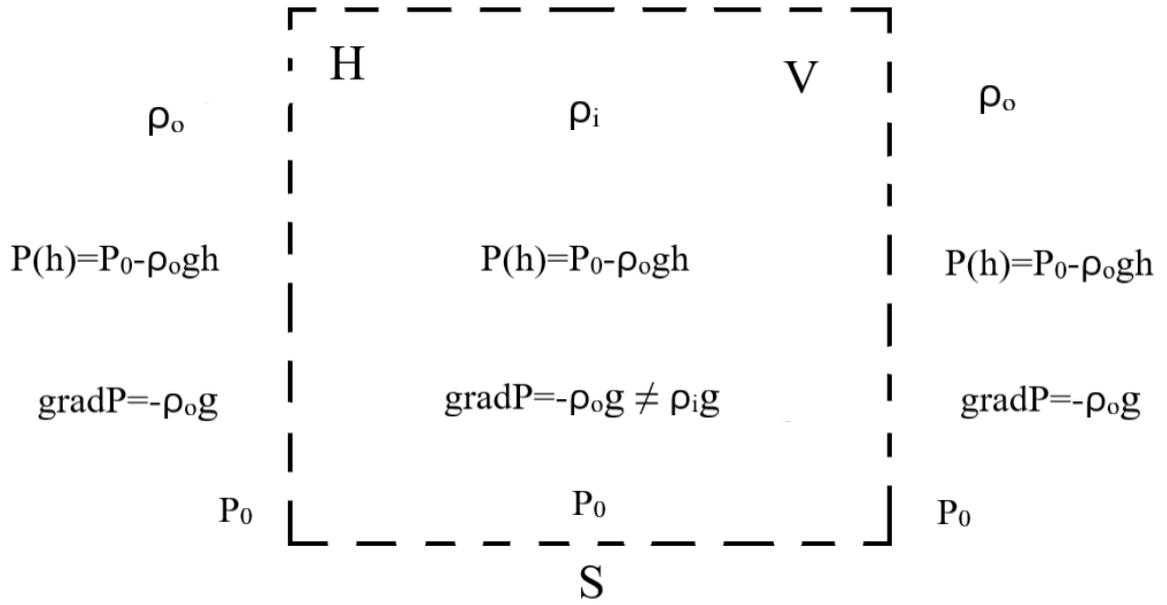

Fig.3. Formation of buoyancy force in gases. Volume V with height H and base area S is highlighted with a dotted line (V=HS). P(h) pressure in the surrounding space. $P_0$ is the pressure on the lower face of the volume, $\rho_0$ is the density of gas in the surrounding space, $\rho_i$ is the density of gas inside the volume, g is the acceleration of gravity, h is the height above the level of the lower face of the volume.

The process of formation of the buoyancy force in gases can be analyzed using the example of volume V with a base value of S and a height of H. The volume of a gas with a density of $\rho_i$ is located in a space that is filled with a gas with a density of $\rho_0$. To simplify, the vertical density is considered constant. The pressure P(h) in the surrounding space is equal to

$$P(h) = P_0 - \rho_0 g h \qquad (1)$$

Where $P_0$ is the pressure on the lower face of the volume, $\rho_0$ is the density of the gas, g is the acceleration of gravity, h is the height above the level of the lower face of the volume. Volume V is highlighted with a dotted line in Fig.3. The horizontal pressure both in the surrounding space and inside the volume at any height is constant, since the process in which the transients of pressure distribution have completed is being considered. The external pressure creates a vertical pressure gradient in volume V equal to

$$gradP = -\rho_0 g \qquad (2)$$

The magnitude of this gradient differs from its own hydrostatic gradient in volume V, which is formed due to its density $\rho_i$

$$-\rho_i g \neq -\rho_0 g \qquad (3)$$

The difference between the external "imposed" gradient and its own gradient forms a volumetric non-hydrostatic addition to the pressure in volume V. This forms a mass force F, which acts on the entire volume V.

$$F = (\rho_o - \rho_i)gHS \qquad (4)$$

Numerically, the mass force F in the ratio (4) corresponds to the buoyancy force. The ratio for the specific buoyancy force can be obtained using the equation of state of the gas

$$P = R_C \rho T \quad (5)$$

Where $R_c$ is the specific gas constant, $\rho$ is the density of the gas, T is the temperature °K. If the temperature in the volume is greater by dT, then for the specific buoyancy force $F_{sb}$, taking into account the equation of state, we can write

$$F_{sb} = \rho_i g \frac{dT}{T} \approx \rho_o g \frac{dT}{T} \quad (6)$$

Where $\rho_i$ is the density of the gas, g is the acceleration of gravity, T is the ambient temperature. It should be especially noted that dT is a horizontal temperature change. The vertical temperature difference does not contribute to the buoyancy force.

### 5. The criterion of stability of the atmosphere.

To predict the formation of convective clouds, a number of different physical and statistical parameters (indices, predictors) of instability have been proposed [Doswell 2006, Holton 2004], in which convective processes are described not directly, but indirectly.

One of the widely used indices is the method of forecasting instability based on available potential energy. The potential Available energy of instability (Convective Available Potential Energy, CARE) is the work that an air particle can potentially do during adiabatic ascent. CAPE is calculated using the equation

$$CAPE = \int_{LFC}^{EL} g \left( \frac{T_p - T_e}{T_e} \right) dz \quad (7)$$

Where EL – Neutral buoyancy level, LFC – Free convection level, g is the acceleration of gravity, $T_p$ is the temperature of the particle, $T_e$ is the temperature of the surrounding air mass

Along with the SAPE, the convection inhibition energy (CIN) is also calculated

$$CIN = \int_0^{LFC} g \left( \frac{T_p - T_e}{T_e} \right) dz \quad (8)$$

LFC – Free convection level

This index determines the magnitude of the energy of the convection counteraction forces.

The basis for these criteria is the change in buoyancy depending on the temperature stratification of the atmosphere. The justification of the ratios for CAPE and CIN is based on dynamic processes with a small displacement of an air particle (Holton 2004).

The analysis for the buoyancy force in a static state made in this work allows us to obtain similar ratios. The air temperature in the particle rising from the level of neutral buoyancy will vary along the dry adiabate, so its potential temperature will be equal to the initial one. As a result, the ratio (6) divided by the density p corresponds to the integrand in the definition of both CAPE and CIN.

### 6. Conclusion.

Due to the absence of rigid boundaries inside the gas, the horizontal pressure gradient cannot exist in a static state, since the pressure equalizes with sound velocity. In this case, the pressure inside the allocated volume with a different density is equal to the external pressure and the intra-mass buoyancy force, according to its definition, is formally zero. The observed force of intra-mass buoyancy arises due to the difference in vertical pressure gradients in media with different densities. The buoyancy force in this case is volumetric, and its value is determined by known ratios. The magnitude of the buoyancy force in the case of a volume located next to a wall requires separate consideration (Lima 2012).

## Acknowledgments

The author thanks colleagues from the Moscow Institute of Physics and Technology for useful discussions during the work.